\documentstyle[prd,aps,multicol]{revtex}

\begin{document}
\draft
\preprint{IAG/123}
\title{Evolution of a primordial black hole population}
\author{ P.S.Cust\'odio and J.E.Horvath  }
\address{ Instituto Astron\^omico e Geof\'{\i}sico, Universidade
de S\~ao Paulo,\\
Av. Miguel St\'efano 4200, \'Agua Funda,
04301-904, S\~ao Paulo, Brasil. }
\date{\today}
\maketitle
\begin{abstract}
We reconsider in this work the effects of an energy absorption term in the 
evolution of primordial black holes (hereafter PBHs)
 in the several epochs of the Universe. A critical mass is 
introduced as a boundary between the accreting 
and evaporating regimes of the PBHs.  
We show that the growth of PBHs is negligible in the Radiation-dominated 
Era due to scarcity of energy density supply from the expanding background, 
in agreement with a previous analysis by Carr and Hawking, 
but that nevertheless the absorption term 
is large enough for black holes above the critical mass to preclude
their evaporation until the universe has cooled sufficiently.
The effects of PBH motion are also discussed: the Doppler effect may 
give rise to energy accretion in black-holes 
with large peculiar motions relative to background.
We discuss how cosmological constraints are modified by 
the introduction of the critical 
mass since that PBHs above it do not 
disturb the CMBR. We show that there is a large range of admissible 
masses for PBHs above the critical mass but well below the cosmological horizon.
Finally we outline a minimal kinetic formalism,  
solved in some limiting cases, to 
deal with more complicated cases of PBH populations.

\end{abstract}

\begin{multicols}{2}
\section{Introduction}

The possibility that PBHs may form in the early universe is 
allowed by many theoretical models. 
Several processes might lead to their formation; 
for instance, the collapse of primordial 
density fluctuations could have an 
important role in their creation. A detailed review of these mechanisms and 
of their effectiveness for creating 
PBHs can be found in Ref.[1]. 
Since the density to which matter
must be compressed in order to create a black hole grows with mass, the 
powerful pressures counteracting the compression must be taken into account. 
As a result, black holes with mass $m \, << \, M_{\odot}$ 
can not form in the contemporary 
universe. Zel'dovich and Novikov [2] and Hawking [3] 
hypothesized that black-holes 
could have been produced at some early stages of the universe. 

PBHs have been studied intensively since the discovery of 
Hawking's quantum evaporation, which scales as the inverse of the squared PBH 
mass. Zel'dovich and coworkers 
analyzed the question of the environment effect on the 
individual PBH evolution.
Here, we shall return to this point
and discuss 
the PBHs behavior taking into account the energy content of medium, their 
proper motions and the causal 
constraints due to existence of a cosmological horizon. Since the abundance of 
these objects may be probed experimentally, the consideration of the actual 
evaporation history is in order. We discuss the regimes of evaporation and 
accretion in Section 2. Section 3 presents an asessment of the kinematical 
effects on the PBHs. Section 4 is devoted to a brief discussion of a 
more general formalism to deal with a generic PBH population. The conclusions 
are presented in Section 5. 

\section{Evaporating and non-evaporating regimes of PBHs}

\subsection{The critical mass}

The equation that describes the evaporation of black holes due to the 
Hawking radiation is [3]

\begin{equation}
{dm \over {dt}} \, = \, - {A(m) \over{m^{2}}} \, \, ;
\end{equation}

where $A(m)$ counts for relativistic particles emitted from a black-hole with 
mass $m$. Numerically, 
$A(m) \, = \, 5.3 \, \times \, 10^{25} \, g^{3} \, s^{-1}$ 
for black holes 
with masses $> \, 10^{17} \, g$ 
and  $A(m) \, \geq \, 7.8 \, \times \, 10^{26} \, g^{3} \, s^{-1}$ 
for black holes with masses $\leq \, 10^{15} \, g$ [4].
However, this expression is strictly valid for 
black-holes immersed in a vacuum, since it 
does not take into account the energy 
content of background. The next step is to consider a term which 
explicitely contains the effective 
energy density and the appropriate cross section for gravitational capture 
of relativistic particles of this background. These effects have been 
first considered by Zel'dovich and Novikov [2] 
(see also Ref.[5] for a discussion) 
in the past. We reexamine the physical picture emerging 
from the introduction of an absorption term in the present work.
We wish to take into account relativistic effects in this description, 
so that an effective mass density can be defined as

\begin{equation}
\rho_{eff} \, = \, \rho  \, + \, 3 p(\rho)/c^{2} \, \, .
\end{equation}

If the environment is assumed to be isotropic and homogeneous, the 
cross-section for relativistic particle absorption is proportional 
to the square of the mass

\begin{equation}
\sigma (m) \, \simeq \, 16 \, \pi {G^{2} m^{2} \over {c^{4}}} \, \, .
\end{equation}

This expression should be interpreted as an order-of-magnitude estimation 
for the cross-section, as is well-known [2] the corrected relativistic 
expression is in fact larger by a factor $1.68$ and we shall neglect 
this small difference in the rest of the paper. 
However, it can be checked the the results depend weakly on the precise 
numerical coefficient of $\sigma (m)$ (see for example Ref.[2]). 
The amount of mass-energy absorbed by the black-hole 
in a time interval $d t$ is therefore

\begin{equation}
{dm \over{dt}} \, = \, {16 \, \pi \, G^{2} \over {c^{3}}} \, m^{2}
\rho_{eff} \,  .
\end{equation}

Then, a more complete differential equation for a description of the PBHs  
mass variation is

\begin{equation}
{dm \over{dt}} \, = -{A(m) \over{m^{2}}} \, + \, 
{16 \, \pi \, G^{2} \over{c^{3}}}
 m^{2} \rho_{eff} \, .
\end{equation}

where the background equation of state enters through $\rho_{eff}$. 
A critical mass is 
obtained by setting eq.(5) equal to zero and reads

\begin{equation}
m_{c} \, = \, {\biggl[ {A_{max} c^{3} \over {16 \pi G^{2}  
\rho_{eff}}} \biggr]}^{1/4} ; 
\end{equation}

where we have chosen arbitrarily $A \, = \, A_{max}$, the maximum 
possible value. Any 
astrophysical or PBH whose mass stays above $m_{c}$ at 
a given cosmological 
time is able to accrete more
mass-energy than the amount that evaporates via quantum effects.
We will now discuss the behavior of this 
parameter during the Radiation  
and the Matter-dominated Eras.

\subsection{Evolution in the Radiation-dominated Era}

In this epoch, the relationship between $\rho_{rad}$ and $p$ is given by 
$p \, = \, \rho_{rad} c^{2}/3$, then the critical mass becomes

\begin{equation}
m_{c} \, = \, {\biggl( {A_{max} c^{3} \over{32 \pi G^{2} \rho_{rad}}} \biggr)}^
{1/4} .
\end{equation}

Numerically, we can express $m_{c}$ in terms of the background 
temperature (in $MeV$) and the relativistic degrees 
of freedom of the background particles $G(T)$ through the 
expression $\rho_{rad} \, = {\pi^{2} \over{30 \, c^{2}}} \, G(T) \, T^{4}$ 
yielding

\begin{eqnarray}
m_{c} \, \approx  \, 2.5  \, \times \, 10^{16} \, 
G(T)^{-1/4} \, {\biggl( {T \over{MeV}} \biggr)}^{-1} 
\nonumber\\ 
\times {\biggl( {A_{max} \over{7.8 \, \times \, 10^{26} \, g^{3} \, s^{-1}}} 
\biggr)}^{1/4} \, g \, \, ;
\end{eqnarray}

or, using the well-known time-temperature relation 
$ T \, = \, 1.55 \times \, G(T)^{-1/4} (t/1 \, s)^{-1/2} \, MeV$

\begin{equation}
m_{c} \, \simeq \, 4 \, \times \, 
10^{17} \, {\biggl( {t \over{1 \, s}} \biggr)}^{1/2} \, 
{\biggl( {A_{max} \over{7.8 \, \times \, 10^{26} \, g^{3} \, s^{-1}}} 
\biggr)}^{1/4} g \,  \, ,
\end{equation}

and we have assumed the validity of the Standard Model, for which one 
gets a maximum value $G_{max}(T \, = \, 300 \, GeV) \, \simeq \, 106.7$ 
associated to the maximum number of particle degrees of freedom; 
and imposed a 
maximum value of $A_{max} \, \equiv \, A( T_{BH} \, \geq \, 300 \, GeV)$ 
when the black hole emits all the known elementary particles.

A plot of $m_{c}$ for this Era is given in Fig.1.
The mass contained in the cosmological horizon defined as 
$m_{h} \, = \, V_{geo}(t) \rho_{rad}(t)$, where 
$V_{geo} \, \approx \, 4 \pi r_{h}^{3}$ is much 
larger than $m_{c}$ except at the Planck epoch. For a 
model with $\Omega \, = \, 1$, $r_{h}$ is given by 
the usual $r_{h} \, \equiv \, R(t) \, \int_{0}^{t} c \, dt'/R(t')$. 

%EDITOR: please put Figure 1 here !

Since in the Radiation-dominated Era $m_{h} \, \approx \, 4 \times 10^{38} 
(t / 1 \, s ) \, g$, 
then sub-horizon PBHs can form and 
stay above the critical mass. By comparing $m_{h}$ and $m_{c}$ we find
a large range of admissible PBH masses in the accretion regime. 

It should be noted that the critical mass expression displayed above is 
strictly valid if the PBHs are scarce
relative to radiation, or $\rho_{PBH} \, \ll \, \rho_{rad}$. 
If this is 
not the case, the evaporation of these objects may inject energy 
into the medium, which in turn modifies 
the critical mass. A set of equations for a more complete description of the 
latter situation is given by

\begin{equation}
H^{2} \, = \, {8 \pi G \over{3}} \, {\bigl( \rho_{rad} \, + \, 
\rho_{pbh} \bigr)} \, - \, {k \over{R^{2}}}
\end{equation}

\begin{equation}
{dm \over{dt}} \, = \, -{A(m) \over{m^{2}}} \, + \, 
{32 \, \pi \, G^{2} \over{c^{3}}}  \rho_{rad} m^{2}
\end{equation}

\begin{equation}
4 H \rho_{rad} \, + \, 3 H \rho_{pbh} \, + \, \dot{\rho}_{rad} \, + \, 
\dot{\rho}_{pbh} \, = \, {{\dot Q}(m) \over{R^{3} \, c^{2}}}  \, \, .
\end{equation}

Where ${\dot Q}(m)$ is the input power due to evaporating black-holes of 
a given mass, which obviously vanishes for $ m \, > \, m_{c}$. We 
consider a flat model of universe 
without cosmological constant. A simple algebraic manipulation of 
these equations yields

\begin{equation}
m_{c} \, \propto \, {\bigl( 1 \, + \, {\rho_{pbh} \over{\rho_{rad}}} \bigr)}^
{1/4} \, H^{-1/2} .
\end{equation}

This expression reduces to the former when $\rho_{pbh}/\rho_{rad} \, \ll \, 1$, 
as it should. 

In the general case the initial spectrum can be quite 
complicated [6]. In the specific case of scale-invariant 
density perturbations (satisfying the condition 
$1/3 \, < \, \Delta \rho /\rho \, < 1$) 
the formation of a continuous 
PBH mass spectrum is expected [7].
If the mass spectrum of PBHs favors light objects, 
a large fraction of 
the population will be evaporating 
in the Radiation-dominated Era, moreover, 
if $\rho_{pbh} \, \approx \, \rho_{rad}$, 
we may get a significative reheating 
of the background (see, for example Ref.[8]) and the critical mass 
would be diminished in turn. Generally speaking 
the radiation from the evaporating subpopulation 
could feed for the non-evaporating massive black-holes above $m_{c}$. 
Consequently, the spectrum 
would be distorted in a non-trivial way.

\subsection{Evolution in the Matter-dominated Era}

 Following the same reasoning than in the former point 
it is straightforward to obtain the 
evolution of critical mass in the Matter-dominated Era. 
We know that the scale-factor 
evolves as $R(t) \, \propto \, t^{2/3}$ and the 
background temperature scale as $T(t) \, \propto \, R(t)^{-1}$, 
then, using again the above formulae, the 
critical mass for this epoch is found to be

\begin{equation}
m_{c} \, \approx \, 7 \times \, 10^{25} 
{\biggl( {t \over{t_{H}}} \biggr)}^{2/3} \, g \, \, .
\end{equation}

Where $t_{H}$ denotes the Hubble time of expansion. 
This result obviously applies to 
astrophysical black-holes as well, as all of the latter 
objects are far above the present critical mass and therefore 
will be accreting photons for a long time.
One may wonder about the actual amount of accretion/mass gain [2,5]. 
We can check that the rate of 
growth of the critical mass $m_{c}$ is in fact much larger than 
the rate of growth of any particular 
black-hole from a reasonable formed spectrum. 
In this case, we can set $m (t_{H}) \, = \, m_{0} \, \approx \, 
constant$ and obtain the approximate equality

\begin{equation}
m_{0} \, \simeq \, m_{c}(t_{max}) \, \, ;
\end{equation}

which is the {\it minimal} mass of black holes that survive in a closed 
universe
(here, $t_{max} \, = \, (\pi/2 \, H_{0}) \, \sqrt \Omega \, 
\approx \, \pi t_{H}/2$ 
is the cosmological time at maximal expansion of a universe with 
$\Omega \, \geq \, 1$). Numerically we 
obtain $m_{0} \, \approx \, 10^{26} \, g$ 
for this minimal mass. 
With this result, we conclude that all astrophysical black-holes should 
survive in closed universes. From the 
maximal expansion, the subsequent contraction phase will be raising 
the temperature of the background and 
these black-holes begin to grow even faster than in the previously 
expanding phase.

\subsection{Absence of relativistic growth of PBHs} 

Now, we will proceed to develop an argument to show that 
similarity solutions do not exist in the Radiation-dominated 
Era. Our arguments are based only on the nature of the equation of 
state for this epoch and the existence of a 
cosmological horizon. By calculating the ratio

\begin{equation}
\Gamma \, \equiv {{\mid {dm_{abs}/{dt}} \mid} 
\over{{\mid {dm_{evap}/dt}\mid}}} \, 
= \, {\biggl( {m \over{m_{c}}} \biggr)}^{4} .
\end{equation}

we imediately conclude that a black-hole with  $m \, \gg \, m_{c}$ absorbs  
a net positive amount of radiation. 
Such an object may be easily created with a mass 
between $m_{c}$ and $m_{h}$ for all 
cosmological times (see eq.9).
The question is : is a relativistic growth of 
these objects possible?
For those black-holes with $m \, \gg \, m_{c}$, 
the evaporation term can be neglected and we get simply

\begin{equation}
{dm \over{dt}} \, = \, {32 \pi G^{2} \over{c^{3}}} \, \rho_{rad} \, m^{2} \, .
\end{equation}

On the other hand, in the
relativistic growth regime the mass increase satisfies 

\begin{equation}
{dm \over{dt}} \, = \, {c^{3} \over{2G}} \, \, .
\end{equation} 

The condition eq.(18) is necessary and sufficient if we ignore the 
transitory existence  
of a vacuum around the black-hole immediately after its formation. 
Comparing eqs.(17) and (18) we get a 
lower bound to the ambient density 
for a relativistic growth of PBHs to be possible, namely

\begin{equation}
\rho_{rad} \, \geq \, 1.2 \, \times \, 10^{52} 
{\biggl( {m \over{10^{15} \, g}} \biggr)}^{-2} \, g \, cm^{-3} \, \, ;
\end{equation}

Relating the density to the cosmic time we find that the 
universe can be dense enough to feed a relativistic growth of a PBH with a 
given mass for times earlier than

\begin{equation}
t_{rel}(m) \, \leq \, 6 \, \times \, 10^{-24} {\biggl( {m \over{10^{15} \, g}} 
\biggr)} \, s \, \, .
\end{equation}

However, the very existence of a cosmological horizon imposes the condition 

\begin{equation}
t_{causal} \, \geq \, 5 \, \times 10^{-24} {\biggl( {m \over{10^{15} \, g}} 
\biggr)} \, s \, \, ;
\end{equation}

for any causal formation of a PBH of a given mass $m$.
Comparing eqs.(20) and (21) we obtain a small time interval for considerable 
accretion to occur, if at all.
The subsequent expansion quickly diminishes the absorption term and the 
absorbed mass-energy becomes a very tiny 
fraction of the original mass. Then, we conclude that 
causality considerations 
leave no room for a substantial growth of PBHs due to an 
increasingly rarefied 
background. We note that $dm/dt \, \rightarrow \, c^{3}/2 G$ only if
$m \, \rightarrow \, m_{h}$.
Since we expect very massive black-holes 
to be rare in realistic 
mass-spectra, then, the above effects 
rule out a substantial mass gain for those black-holes 
satisfying $m_{c} \, < \, m \, \ll \, m_{h}$ and the mass increase remains 
a moderate one [5]. Another way to understand this effect is to integrate 
eq.(17) to give 
$m(t) \, \simeq \, m_{0} \, (1 \, + \, 3.5 \times
10^{-39}[m_{0}/g][t_{0}/s]^{-1})$; where $m_{0}$ is the initial mass and
$t_{0}$ the formation time of the black-hole in the Radiation Era. Since the 
gain for black-holes inside 
$m_{c} \, < \, m_{0} \, \ll \, m_{h}$ is very small, we again 
conclude that a relativistic growth is precluded.

It is important to notice that, since the critical mass itself grows 
much faster than the mass of those 
black-holes that stay above it, we can estimate the time 
a given black-hole (formed at $t_{F}$ with 
original mass $m_{0} \, \equiv \, m(t_{F})$, above $m_{c}$ by hypothesis) 
spent in the accreting regime inside the Radiation-dominated Era. 
This period is denoted by $t_{c}(m_{0})$
and is given by condition $m(t_{c}) \, = \, m_{c}(t_{c})$.
Since $\Gamma \, = \, (m/m_{c})^{4}$, we may ignore the Hawking term when 
$m_{0} \, \gg \, m_{c}(t_{F})$, and invert an analytical solution of 
the condition $m(t_{c}) \, = \, m_{c}(t_{c})$, yielding

\begin{equation}
t_{c} \, \approx \, 400 \, 
{\biggl( {m_{0} \over{10^{15 \, g}}} \biggr)}^{2} \, s \, \, ; 
\end{equation}

in the Radiation-dominated Era and 
$t_{c} \, \simeq \, 2 \, \times \, 10^{10} \, t_{H} \, 
(m/m_{\odot})^{3/2}$ in the Matter-dominated Era.
We may compare this value to the evaporation time due to quantum
effects alone $t_{Hawking} \, \simeq \, 10^{17} \, s (m/10^{15} \, g)^{3}$. 
We conclude that the evaporation time
picks up  a small correction due to the 
background influence, which delays the 
onset of the evaporation regime unless the black hole
is more massive than $10^{23} \, g$. In the latter case the non-evaporating 
regime has a duration comparable to the age of the universe.      

In spite that these considerations show that a
 PBH with, say,  $m \, > \, 3 \times \, 10^{22} \, g$ slowly 
stays accreting energy during the
whole Radiation-Dominated Era, the net gain of energy is very small. For 
example a PBH formed at 
$t \, \simeq \, 1 \, s$ with $m_{0} \, \simeq 10^{20} \, g$ 
gains a mass fraction of $\delta m/m \, \approx \, 10^{-19}$ 
when accreting for $10^{8} \, s$.
We have checked that this feature holds irrespective of the actual 
relationship between $t_{F}$ and $m_{0}$. Then, all those black-holes 
with $m \, >> \, m_{c}$ are better described by $dm/dt \, = \, 0$ provided 
they form a rarified gas. If the transient vacuum formed around the black-hole 
immediately after its birth is considered, accretion is impossible 
before a time of the order of 
$H^{-1}$, and the background radiation would be even more rarified than our 
previous estimates when the steady conditions are re-established. 

\section{Kinematical effects}

Up to now we implicitely assumed that the PBHs are 
at rest relative 
to the thermal background or have negligible 
peculiar velocities, However, the primordial events ocurring in the early 
universe might be responsible for 
imparting substantial 
kinetic energy to black-holes. For instance, if a 
randomly moving black-hole is hit by an 
expanding bubble wall, it is possible that the former can acquire a 
large momentum because of the sudden 
deposition of energy from the relativistically moving bubble wall [6]. 
A black-hole moving with velocity $v_{pec}$
sees the background temperature to have an angular dependence

\begin{equation}
T(\gamma,\theta) \, = \, 
{T \over{\gamma (1 - {v_{pec} \over{c}}\cos \theta)}} \, ; 
\end{equation}

with $T$ being the temperature in the cosmic frame. Due to this 
peculiar motion, the rate of change of mass of the black-hole in 
its rest frame is given by

\begin{equation}
{dm \over{d\tau}} \, = \, - {A(m) \over{m^{2}}} \, + \, 
{32 \pi G^{2} \over{c^{3}}} \, <\rho_{rad}> \, m^{2} \, \, ; 
\end{equation}

where 

\begin{equation}
<\rho_{rad}> \, = \, {1 \over{4\pi}} \, 
\int d \Omega \, \rho_{rad}(T(\gamma, \theta)) \, = \, 
{{(4 \gamma^{2} - 1) \over{3}} \rho_{rad}} \, \, ,
\end{equation}

is the average over the 
angles and $\tau$ is 
the proper time.
After a simple calculation we get, 
using the relation between the proper time 
and the cosmic time $\gamma \, d\tau \, = \, dt$ 

\begin{equation}
{dm \over{dt}} \, = \, - {A(m) \over{\gamma m^{2}}} \, + \, 
{32 \pi G^{2} \over{c^{3}}} 
{\biggl( {{4 \gamma^{2} -1} \over{3 \gamma}} \biggr)} 
\, \rho_{rad} \, m^{2} \, \, .
\end{equation}

A high value of the Lorentz factor 
$\gamma$ enhances the energy-mass absorption term, and at same 
time inhibits the mass loss due to the 
Hawking radiation. These effects may be incorporated in the critical mass 
definition, which becomes 

\begin{equation}
m_{c} (T,\gamma) \, = \, 
{\biggl( {{4 \gamma^{2} - 1} \over{3}} \biggr)}^{-1/4} \,
 \, m_{c}(T) \, \, \, ;
\end{equation}

where $m_{c}(T)$ is defined in eq.(8). 

How do we interpret these results? For an observer 
moving with the black-hole, the 
thermal background is not isotropic and the mass accretion 
can be larger than the 
evaporation. The critical mass is decreased due to the Doppler effect.
From the point of view of an asymptotic observer, comoving with the 
cosmological expansion, both terms are altered: the Hawking radiation 
term is reduced by the Lorentz  factor due 
to the time dilation effect (in the rest frame the 
Hawking radiation depends 
on the rest mass only). The second term is altered due to same Doppler 
effect and also corrected by the Lorentz factor. 
If we have to decide whether a particular 
black-hole is above or below the critical mass, taking into account its 
possible peculiar motion, we must resort to
the rest frame or the comoving frame. 
Any other frame is pertubed by the proper motion of the observer 
and the results are altered. 
For a black-hole with rest mass $m$ we can calculate the 
minimum peculiar motion 
necessary to ensure that the absorption term 
overcomes the Hawking radiation term. Thus, we can define the threshold 
Lorentz factor

\begin{equation}
\gamma_{th} \, = \, {1 \over{2}} \, {\sqrt {1 + 3 / \Gamma}} \, \, . 
\end{equation}

A black-hole which is above this bound (see Fig.2) will 
accrete mass, even if $m \, < \, m_{c}$, and may stay in this condition 
for a long time depending on the actual evolution of the $v_{pec}$ field.

%EDITOR: please put Figure 2 here !

\section{An outline of a kinetic theory}

In order to study the detailed evolution of a PBH population one 
must go beyond the thermodynamical description and set up a 
formalism capable of describing the variety of physical effects 
existing in the early universe. Such an analysis 
may be achieved through a set of differential equations for the 
mass distribution 
function of black-holes $f$, the cosmological expansion, 
and the equation for mass variation. The 
initial conditions must also be specified (for example, the 
fluctuation pattern that originates a given initial mass spectrum). 
A minimal kinetic formalism is given by

\begin{equation}
{\partial f \over{\partial t}} \, + \, {d \mu \over{d t}} 
{\partial f \over{\partial \mu}} \, + \, 
{d^{2} \mu \over{d t^{2}}} 
{\partial f \over{\partial \dot{\mu}}} \, - \, H \beta 
{\partial f \over{\partial \beta}} \, = \, 0
\end{equation}

\begin{equation}
H^{2} \, = \, {8 \pi G \over{3}} {( \rho_{rad} \, + \, \rho_{pbh} )} \, - 
\, {k \over{R^{2}}}
\end{equation}

\begin{equation}
{d\mu \over{d t}} \, = \, - {A(m) \over{ m_{Pl}^{3} \mu^{2}}} \, + 
\, {32 \, \pi \, G^{2} \, m_{Pl} \over{c^{3}}}  
\rho_{rad} \mu^{2}  
\end{equation}

\begin{equation}
4 \, H \, \rho_{rad} \, + \, 3 \, H \, \rho_{pbh} \, + \, 
\dot{\rho}_{rad} \, + \, 
\dot{\rho}_{pbh} \, = \, 
{\dot{Q}(m) \over{R^{3} \, c^{2}}} \, \, ;
\end{equation}

where $\mu$ is the dimensionless 
PBH mass scaled to the Planck mass $m_{Pl}$, 
$f$ is the distribution 
function defined on an extended 
phase-space including the mass.  
With these definitions we have $\rho_{pbh}(m,t) \, \equiv \, m \, n(m,t)$. 
Given some initial conditions 
the solution of this set of equations describes a 
dilute black-hole gas in a non-relativistic regime for a given model of 
the universe.
 Now,we will seek particular solutions for a flat model of universe, a dilute
and sub-dominant PBH-gas, and non-relativistic peculiar motions.
 Since that the $({d \mu \over{d t}})({\partial f \over{\partial \mu}})$ 
term changes sign as a PBH crosses the critical
mass, we split $f$ in two functions defined by

\begin{equation}
f_{+} \, \equiv \, f(\mu \, > \, \mu_{c},t)
\end{equation}

and
\begin{equation}
f_{-} \, \equiv \, f(\mu \, < \, \mu_{c},t)
\end{equation}

Since it can be shown that there is no termodynamical 
equilibrium between the PBH gas and the background radiation, these
objects cannot stay fixed at the critical mass (mainly 
because of the cosmological
expansion). The relevant observables can be calculated if $f_{\pm}$ are known.
When expressed in terms of $f_{\pm}$ the comoving number density of PBHs is 
simply given by

\begin{equation}
n_{\pm} \, = {g(\mu) \over{(2\pi)^{3}}} \, \int {d^{3}p \, f_{\pm} 
(\mu, \beta,t)}
\end{equation}

 with $g(\mu)$ is the statistical weight of the black hole gas and 
the integration runs over all the peculiar
momentum $p \, = \, m_{Pl} \, c \, \mu \, \beta$.
 The geometrical expansion of the universe can be isolated with the choice 

\begin{equation}
f_{\pm}(\mu, \beta,t) \, = \, {\biggl( {R(t) \over{R_{0}}} \biggr)}^{-3} \, 
f^{'}_{\pm}(\mu, \beta,t) \, \, ,
\end{equation} 
 
 where $R_{0}$ is the scale
factor at the formation time for this population (assumed to be unique for 
simplicity). Furthermore, since we have assumed a dilute PBH gas, 
a separation of variables is possible to cast $f_{\pm}$ into the form

\begin{equation}
f_{\pm}(\mu, \beta, t) \, = {\biggl( {R(t) \over{R_{0}}} \biggr)}^{-3} \, 
F_{\pm}(\mu, t) \, J_{\pm}(\beta) \, \, ,
\end{equation}

Within our hypothesis $\rho_{pbh} \, \ll \, \rho_{rad}$ 
(and also that the bulk of the PBH population residing 
between $m_{\ast} \, = \, (3 A / H)^{1/3}$ and $m_{c}$ in order to 
avoid a substantial injection of energy due to low-mass objects), 
the third term in eq.(28) can be neglected and we get

\begin{equation}
{1 \over{F_{\pm}}} \, {\biggl( {\partial F_{\pm} \over{\partial t}} \biggr)} \, 
- \, 3 H \, + \, 
{\dot{\mu} \over{F_{\pm}}} 
{\biggl( {\partial F_{\pm} \over{\partial \mu}} \biggr)} \, 
- \, {H \, \beta \over{J_{\pm}}} \, 
{\biggl( {d J_{\pm} \over{d \beta}} \biggr)} \, = \, 0 \, .
\end{equation}
 
The quantity $d \beta \, \beta^{2} \, J_{\pm}(\beta)$ is the comoving number
of black holes with peculiar velocities belonging to the interval 
$(\beta, \beta \, + \, d \beta)$.
We will show below that the minimal kinetic formalism is consistent 
with the concepts deduced in the previous sections, in particular the 
negligible mass accretion for black
holes described by $f_{+}$.
 
The number density of the evaporating $n_{-}$ and non-evaporating 
$n_{+}$ populations 
is given in terms of the new variables as

\begin{equation}
n_{\pm}(\mu, t) \, = \, {g(\mu) (m_{Pl} c)^{3} 
\over{2 \pi^{2}}} \, \mu^{3} \, F_{\pm} \, 
{\int_{0}^{1} d \beta \, \beta^{2} \, J_{\pm}(\beta)} \, ;
\end{equation}

respectively. The kinematical effects discussed in section III can be taken 
into account by replacing the upper limit of the integration over 
$J_{-}$ by $\beta_{th}(\mu) \,  = \, \surd (1 \, - \, 1/\gamma_{th}^{2})$, 
where $\gamma_{th}$ is given by eq.(27). We shall neglect this effect in 
the following.       

We finally arrive to the expressions of the comoving mass density for each 
subpopulation as

\begin{equation}
\rho(\mu \, > \, \mu_{c}, t ) \, = 
\, {g(\mu) \, m_{Pl}^{4} \, c^{3} \, I_{+} 
\over{2 \pi^{2}}} \, \int^{\mu_{h}(t_{F})}_{\mu_{c}(t)} \, 
d \mu \, \mu^{3} \, F_{+}(\mu, t) \, ,
\end{equation}
 
and

\begin{equation}
\rho(\mu \, < \, \mu_{c}, t ) \, = 
\, {g(\mu) \, m_{Pl}^{4} \, c^{3} \, I_{-} 
\over{2 \pi^{2}}} \, \int^{\mu_{c}(t)}_{0} \, 
d \mu \, \mu^{3} \, F_{-}(\mu, t) \, ;
\end{equation}

 where we have extended the upper limit of the first integration since 
there is no mass gain for black holes with initial mass 
$m_{0} \, \leq \, m_{h}(t_{F})$, as 
discussed in section II.D ; and defined 
$I_{\pm} \, = \, \int_{0}^{1} \, d \beta \, \beta^{2} \, J_{\pm}(\beta)$.
                                
A generic PBH population will show one 
novel feature: if there are some
black holes with $m_{0} \, > \, m_{c}(t_{F})$ 
these objects will be crossing the
critical mass curve and go into the subcritical region. 
The kinetic formalism is able to take into account
this effect in the following way; let us separate 
$\rho(\mu \, < \, \mu_{c})$ as a sum of two
contributions $\rho_{1}$ and $\rho_{2}$, defined by

\begin{equation}
\rho_{1}(\mu \, < \, \mu_{c}) \, \propto \, \int_{0}^{\mu_{\ast}(t)} \, 
d \mu \, \mu^{3} \, F_{-}(\mu , t)
\end{equation}

and

\begin{equation}
\rho_{2}(\mu \, < \, \mu_{c}) \, \propto \, \int_{\mu_{\ast}(t)}^{\mu_{c}(t)}
\, d \mu \, \mu^{3} \, F_{-} (\mu , t)
\end{equation}

where 
$\mu_{\ast}(t) \, \simeq \, \mu_{c}(t_{F}) (1 \, - \, 3 A (t - t_{F})/
m_{Pl}^{3} \mu_{c}(t_{F})^{3})^{1/3}$ is a boundary curve, which 
evidently divides the evaporating black-holes 
into a population that was already born in such condition, and a 
population that entered the evaporation regime after crossing $m_{c}$ in 
a finite time (see Fig.3).  
Since the latter PBHs satisfy $\delta m/m \, \ll \, 1$, 
then their number density is described by 
the usual dilution of non-relativistic objects $n \, = \, n_{0} \,
(R/R_{0})^{-3}$ and therefore we get

\begin{eqnarray}
\rho(\mu \, > \, \mu_{c}, t) \, = \, \int_{m_{c}(t)}^{m_{h}(t_{F})} \, 
d m \, n(m, t)
\nonumber\\
= \, n_{0} {\biggl( {R(t) \over{R_{0}}} \biggr)}^{-3} \, 
(m_{h}(t_{F}) \, - \, m_{c}(t))
\end{eqnarray}

 This density is diluted by two factors, as expected (see Fig.3): the
first is the usual geometrical expansion and the second their
migration into the subcritical regime. It is obvious that 
$\rho(\mu \, > \, \mu_{c} , t \, > \, t_{c}(\mu_{h})) \, = \, 0$, since 
after $t_{c}(\mu_{h})$ no PBHs in the accretion regime remain.
Comparing eqs.(39) and (43) we get that $F_{+} \, \propto \, \mu^{-3} \,
I_{+}^{-1}$. Substituting $F_{+}$ into eq.(37) and solving for $J_{+}$ 
we obtain

\begin{equation}
J_{+}(\beta) \, = \, J_{0} \, \beta^{-G_{+}(t)}
\end{equation}

where the exponent is $G_{+}(t) \, = \, 3(1 \, + \, 2 {\dot \mu}/\mu \, H)$ 
(with ${\dot \mu}/\mu \, H \, \ll \, 1$) and $J_{0}$ is defined by the 
initial conditions.

%EDITOR: please put Figure 3 here ! 

On the other hand, the subcritical population is much more complicated 
than the supercritical one. 
To simplify the solutions we assume that the source term in 
the last eq.(31) may be written as 
${\dot Q(m)} / R^{3}(t) \, c^{2} \, \sim \, \rho(m) / \tau_{evap}(m)$, 
where $\tau_{evap} \, = \, (m_{0}^{3} \, - \, m^{3}) / 3 \, A(m)$ 
is the timescale for a black-hole to evaporate from $m_{0}$ to $m$, and 
get the following solution

\begin{eqnarray}
F_{-} \, \propto \, {\mu^{-3} \over{I_{-}}} 
(\exp(-t \Gamma_{1}) \, \Theta(\mu \, - \, \mu_{\ast}) \, + 
\nonumber\\
+ \, \exp(-t \Gamma_{2}) \, \Theta(\mu_{\ast} \, - \, \mu)) 
\end{eqnarray} 

where $\Theta$ is the step function and $1 / \Gamma_{1} , 1 / \Gamma_{2}$ 
are mass-dependent timescales defined by

\begin{equation}
\Gamma_{1} \, = \, {\biggl( {3 \, A(m) \, 
\over{m_{Pl}^{3}(\mu_{0}^{3} \, - \, \mu^{3}}}\biggr)} \; \; \; \; 
\mu_{0} \, \leq \, \mu_{c}(t_{F})
\end{equation}

and 

\begin{equation}
\Gamma_{2} \, = \, {1 \over{t_{c}(\mu_{0})}} \, 
{\biggl( 1 \, + \, {(\mu_{0}^{3} \, - \, \mu^{3}) \, m_{Pl}^{3} 
\over{3 \, A(m) t_{c}(\mu_{0})}} \biggr)}^{-1} 
\; \; \, \mu_{0} \, \geq \, \mu_{c}(t_{F})
\end{equation}

Therefore, substituting $F_{-}$ in eq.(37) we obtain the solution 

\begin{equation}
J_{-}(\beta) \, = \, J_{0} \, \beta^{-G_{-}(t)} \, ;
\end{equation}

and the exponent is given by

\begin{equation}
G_{-}(t) \, = \, 3 \, {\biggl( 1 \, + \, 
{2 {\dot \mu} \over{\mu \, H \,}}  \, - \, 
{{\dot X}  \over{3 \, H \, X}} \, - \, {{\dot \mu} 
\over{3 \, X \, H}} {\partial X \over{\partial \mu}} \biggr)} \, ;
\end{equation} 

where ${\dot X}$, etc. are time derivatives and $X$ is defined as 

\begin{equation}
X \, \equiv \, \exp(-t \Gamma_{1}) \, \Theta(\mu \, - \, \mu_{\ast}) \, + 
\exp(-t \Gamma_{2}) \, \Theta(\mu_{\ast} \, - \, \mu) \, \, .
\end{equation}
                       
The exponential factors $\exp(-t \Gamma_{i})$ explicitly describe the 
evaporation of PBHs.
In fact, $\Gamma_{1}$ controls the rate of evaporation
of PBHs from $\mu_{0}$  to $\mu$, and $\Gamma_{2}$ is the analogous 
timescale, but corrected for the time $t_{c}(\mu_{0})$ when no evaporation 
occurs (see Fig.3).     

An important point (not addressed in this work) is to relate the 
mass-spectrum functions $F_{\pm}(\mu, t)$ to the initial conditions which 
formed the PBH population, for example, large primordial fluctuations  
imprinted in the radiation field. A handful of works in the literature 
address this problem, i.e. Ref.[8]. Carr's mass spectrum corresponds  
to the particular choice $F \, \equiv \, \mu^{n - 3} \, N(\mu, t)$ (where 
$N(\mu, t) \, = \, A \, \mu^{-n}$ and $A \, = \, constant$) 
in the presented formalism. 
The accreting subpopulation of Carr's spectrum $F_{+}$ is not specially 
affected by our considerations because that each of that 
objects remain near the formation mass. However, the evaporating part 
$F_{-}$ depends quite  
strongly on time, and therefore the initial spectrum is modified with 
yet unknown consequences. Incidentally, eqs.(45) and (49), show  
that the the PBH population does not enter into equilibrium with 
the ambient radiation in this approximation.

As an application of the above results we may estimate the maximum 
allowed number density (at a given formation time) consistent with the 
observed expansion and the present bound $\Omega_{pbh} \, < \, 10^{-8}$ [9].
If we consider a power-law mass spectrum of the type 
$N \, \propto \, \mu^{-n}$  
and a "burst" of PBHs formation at $t \, = \, t_{F}$ in the 
Ratiation-dominated Era, the above expressions can be manipulated to give

\begin{eqnarray}
n (t_{F}) \, \, \, \, < \, \, \, \; 3.5 \, \times \, 10^{-42} \, 
\mu_{c}^{(n \, - \, 1)}(t_{F}) \,
\nonumber\\
\; \; \; \; \; \;  \times {\biggl( {t_{F} \over{1 \, s}} \biggr)}^{-5/2}
(1 \, + \delta / 2) \, \Omega_{pbh} \, h^{2} \, cm^{-3}
\end{eqnarray}

with $\delta \, \equiv \, m_{c}(t_{h}) / m_{h}(t_{F}) \, \ll \, 1$.
   
This discussion 
illustrates some features of the formalism, which is of course 
not restricted to Radiation-dominated, flat cosmological models. More 
general conditions and a variety of mass spectra will be presented in 
future works.

\section{Conclusions}

We have reexamined in this work the evaporating and accreting (more 
properly termed as "non-evaporating" because of the smallness of the net 
gain) regimes of PBHs for various epochs in the early universe. We 
have confirmed the existence of a substantially broad mass interval 
$[m_{c}, m_{h}]$ which may have been populated by PBHs which do not  
disturb the CMBR through evaporation if a continuous initial mass 
spectrum is assumed. 
We have also shown that the relativistic growth of PBHs contained 
in $[m_{c}, m_{h}]$ is unlikely by considering the absorption term 
and the evolution of 
the background density , recovering a result due to 
Carr and Hawking [5] which solved the general 
relativistic equations in detail.  

Kinematical effects that may affect the accreting/evaporating fraction 
of PBHs have been also discussed, and a threshold Lorentz factor 
for the accretion to dominate was derived. 
We have also sketched 
how a more complete formalism can be formulated based on 
a kinetic equation for a distribution function on an extended 
phase space that includes the mass as a variable; and solved it in some  
particular cases to make contact with previous works. 
We hope to address in future papers the much more complicated case of 
a population containing both evaporating and non-evaporating 
holes simultaneously, 
and to study cosmological scenarios and observational constraints.

\acknowledgements

We would like to acknowledge the financial support of CAPES and CNPq 
and FAPESP 
agencies (Brazil) through Fellowships awarded to P.S.C. and J.E.H. 
respectively. T.P.Dominici and M.P.Allen are 
acknowledged for technical support and 
advice during this work. Finally we would like to acknowledge the 
criticisms of an anonymous referee, which have helped to improve the 
original draft of this work.

\end{multicols}

\vfill\eject

\section{Figure Captions}

Figure 1. Critical mass $m_{c}$ as a function of the temperature in the 
Radiation-dominated Era. 
The curve splits this plane in two different regimes as 
throughly explained in the text. Note the jumps due to the changing 
degrees of freedom of the background.

\bigskip

Figure 2. Threshold Lorentz factor $\gamma_{th}$ as a fuction of the 
mass ratio $m_{c} / m$. Black holes above the curve accrete matter and 
do not imediately evaporate (in spite of the condition 
$m_{c}/ m \, > \, 1$) due to the $\gamma$ dependence of both the accreting 
and evaporating terms of eq.(26). Those black-holes satisfying 
$m_{c}/m \, < \, 1$ do not evaporate even if their peculiar velocities 
vanish. 

\bigskip

Figure 3. Qualitative graphical representation of the evolution of 
a generic PBH population in the $\mu - t$ plane. The crossing-time 
$\tau_{c} (\mu_{0})$ for a black hole with initial mass $\mu_{0}$ is 
depicted, together with the horizon mass $\mu_{h}$ and critical mass 
$\mu_{c}$ defining the different regimes in the Radiation-dominated Era 
(see the text for details).

\end{document}